\documentclass[a4paper,11pt]{article}

\usepackage{pos}
\usepackage{lipsum} 
\usepackage{caption}
\usepackage{subcaption}
\usepackage{tabularray}
\usepackage{xcolor}
\usepackage{siunitx}
\usepackage{csquotes}
\usepackage{xfrac}
\usepackage{placeins}
\usepackage[nameinlink]{cleveref}

\makeatletter
\newcommand{\rcite}[1]{%
    \begingroup%
    \hypersetup{hidelinks}%
    \hyper@natlinkstart{#1}%
    Reference~\cite{#1}%
    \hyper@natlinkend%
    \endgroup%
}
\makeatother

\UseTblrLibrary{booktabs}
\UseTblrLibrary{siunitx}

\title{Measuring the Neutrino Flux in Segments along the Galactic Plane with IceCube}

\ShortTitle{IceCube Segmented Galactic Neutrino Flux}

\author{The IceCube Collaboration \\{\normalsize \normalfont(a complete list of authors can be found at the end of the proceedings)}\\}

\emailAdd{ludwig.neste@fysik.su.se}
\emailAdd{mhuennefeld@icecube.wisc.edu}
\emailAdd{cfinley@fysik.su.se}

\abstract{

Gamma-ray emission from the plane of the Milky Way is understood as partly originating from the interaction of cosmic rays with the interstellar medium. The same interaction is expected to produce a corresponding flux of neutrinos. In 2023, IceCube reported the first observation of this galactic neutrino flux at 4.5$\sigma$ confidence level. 
The analysis relied on neutrino flux predictions – based on gamma ray observations – to model the expected neutrino emission from the galactic plane. 
Three signal hypotheses describing different possible spatial and energy distributions were tested, where the single free parameter in each test was the normalization of the neutrino flux.
We present first results of an analysis that can improve the characterization of Galactic neutrino emission by dividing the galactic plane into segments in galactic longitude.
An unbinned maximum-likelihood analysis is used that can fit the spectral index and the flux normalization separately in each segment. 
While gamma ray telescopes can not differentiate between hadronic and leptonic emission, neutrino production must come 
from hadronic processes.
Measuring a spectral index can further help to understand the contribution of unresolved neutrino sources inside the galactic plane.
This work uses a full-sky cascade dataset and provides model-independent insight into the variation of the neutrino flux and energy distribution from different regions of the galactic plane.

\vspace{4mm}

{\bfseries Corresponding authors:}
Ludwig Neste$^{1*}$, 
Mirco Hünnefeld$^{2}$, 
Chad Finley$^{1}$
\\
{$^{1}$ \itshape Oskar Klein Centre and Dept. of Physics, Stockholm University, SE-10691 Stockholm, Sweden}\\
{$^{2}$ \itshape Dept. of Physics and Wisconsin IceCube Particle Astrophysics Center, University of Wisconsin{\textemdash}Madison, Madison, WI 53706, USA}\\[4mm]
$^*$ Presenter
}

\ConferenceLogo{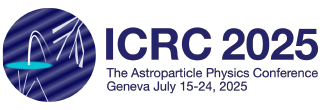}

\FullConference{39th International Cosmic Ray Conference (ICRC2025)\\
 15–24 July 2025\\
Geneva, Switzerland\\}

\begin{document}

\maketitle

\section{Neutrino Flux from the Milky Way}\label{sec1}
For many forms of astronomy, the Milky Way is the most prominent source in 
the sky.
It has been observed throughout the electromagnetic spectrum from radio waves to gamma rays.
The gamma rays emitted in the Galactic Plane (GP) are understood to be partially produced 
by cosmic rays interacting with the interstellar medium. 
In these hadronic interactions, charged ($\pi^+, \pi^-$) and neutral ($\pi^0$) pions are 
created approximately at a ratio of 1:1:1. 
The neutral pions decay into two photons ($\pi^0\to \gamma \gamma$), producing a 
diffuse gamma ray emission in the GP, which has been observed beyond 
\qty{100}{\tera\electronvolt} by the LHAASO \cite{LHAASO:2023gne} Tibet AS$\gamma$ \cite{TibetASgamma:2021tpz}
and HAWC \cite{Alfaro_2024}
observatories. 
The charged pions decay into neutrinos ($\pi^+\to \mu^+ \nu_\mu \to e^+\nu_\mu\bar\nu_\mu\nu_e$).
Two years ago, the IceCube collaboration announced the first observation of neutrino emission from the GP
at \num{4.5}$\sigma$ confidence \cite{IceCube:2023ame}.
This discovery-oriented analysis relied on templates, which model the spatial and energy distribution 
of the diffuse neutrino emission using the measurement of diffuse gamma ray emission at lower energies from 
Fermi-LAT \cite{Fermi-LAT:2012edv, Gaggero:2015xza}.
The shape of the spectrum was fixed to the model prediction, and the single free fit parameter was 
the normalization of the spectrum.
We present a new analysis method, which does not rely on a specific model assumption, 
but instead allows fitting a single power law in different regions of the GP.
This method provides a flux normalization and spectral index in each of 
the chosen segments of the GP.

The IceCube observatory consists of 5160 digital optical modules, each equipped with a photomultiplier tube, 
frozen inside a volume of one cubic kilometer of glacial ice 1.5 kilometers below the geographic South Pole.  
When a high-energy neutrino interacts with the ice, it produces charged secondary particles, which 
emit Cherenkov light. 
The Cherenkov light can be used to reconstruct the direction and energy of the primary neutrino.
IceCube records two main event topologies: 
\enquote{tracks} consist of Cherenkov light emitted by a muon along a line.
Muons are created 
in charged-current $W^\pm$ boson interactions between muon neutrinos and atomic nuclei in 
the ice.
Tracks provide the best angular resolution ($<\qty{1}{\degree}$), but suffer 
from a background of atmospheric muons in the southern sky, which occur one million times more frequently than any
neutrino interaction observable by IceCube. 
\enquote{Cascades} occur when any flavor of neutrino interacts inside the detector volume via a neutral current
$Z^0$ boson interaction or electron neutrinos $\nu_e$ and tau neutrinos $\nu_\tau$ interact via any interaction type.
The light from these cascades radiates almost spherically and the angular resolution is thus poorer ($\sim \qty{5}{\degree}$ \cite{IceCube:2024csv}).
Cascades are essential to gain sensitivity in the southern sky, since they are distinguishable from down going atmospheric muons. 

The observation of the GP was enabled by novel machine learning techniques used to select and reconstruct cascades \cite{Abbasi_2021, Huennefeld:2021Jk}, which 
were used to increase the previous cascade selection by a factor of 20 to approximately \num{60000} events.
The same \num{60000} cascade events were used to perform the analysis presented here.

\newpage

\section{Segmenting the Signal Hypothesis}
This work uses the same unbinned maximum likelihood technique as previous IceCube 
neutrino source searches \cite{Braun:2008bg, IceCube:2019cia}.
The likelihood
\begin{equation}
\mathcal{L} ({n_s}, {\gamma})=
\prod_{i=1}^{N} \left[
\frac{n_s}{N} S_i(E_i, \delta_i, \alpha_i, \sigma_i, \gamma)
+ \frac{N-{n_s}}{N} B_i(E_i, \delta_i)
\right]
\label{eq:likelihood_base}
\end{equation}
is evaluated for each event $i$ with its reconstructed quantities energy 
$E_i$,  declination $\delta_i$, right ascension $\alpha_i$ and angular uncertainty $\sigma_i$.
The number of signal neutrinos $n_s$ relative to the total number of neutrino events $N$ and the spectral index of 
the signal probability density function (PDF) $\gamma$ is optimized to maximize the likelihood of the data.
The signal PDFs $S_i$ for each event is constructed using Monte Carlo (MC) simulation, while the 
background PDFs $B_i$ are derived in a data-driven way.
Previously, it was assumed that in a given declination band the data is dominated 
by background events and averaging over right ascension gives a good estimation of the
background PDF.
For the given dataset the contribution from the GP cannot be neglected any longer. 
Thus, the right ascension average of the signal PDF $\tilde S_i$ is subtracted from the averaged data PDF $\tilde D_i$ using 
the method described in \rcite{IceCube:2017trr}, which was also used in the 2023 GP analysis \cite{IceCube:2023ame}.
This gives the signal substracted background PDF 
\begin{equation}
B_i^{\mathrm{SigSub}}(E_i, \delta_i, \gamma) = \frac{N}{N-n_s} \left(\tilde D(\delta_i, E_i) - \frac{n_s}{N} \tilde S_i(\delta_i, E_i, \sigma_i, \gamma)\right).
\label{eq:sigsub}
\end{equation}
Substituting the last term in \Cref{eq:likelihood_base} with \Cref{eq:sigsub} results in the 
Likelihood used in this work.

The analysis presented here introduces additional parameters into the likelihood by \emph{segmenting}
the signal PDF $S_i$ into $M$ parts
\begin{equation}
S_i(E_i, \delta_i, \alpha_i, \sigma_i, w_1, \dots, w_M, \gamma_1, \dots, \gamma_M)
= 
\sum_{k=1}^M
w_k S_i^k(E_i, \delta_i, \alpha_i, \sigma_i, \gamma_k)
\label{eq:segmented_sigpdf}
\end{equation}
where $w_1,\dots, w_M$ are the weights assigned to each segment's PDF $S_i^k$ 
and $\gamma_k$ their spectral indices. 
In order to make $S_i$ a PDF when segmenting, the weights must be normalized $\sum_k w_k = 1$ and positive $w_k \geq 0$, which is 
implemented via a softmax-transformation.
The spatial shape of the segments and their number can be freely chosen. The choice will be called a \enquote{segmentation scheme} here.
In this work we consider spatially disjunct, homogeneously emitting regions, aligning with the galactic
plane in longitude. They are presented in the following sections.
Equivalently, with Equation \eqref{eq:segmented_sigpdf}, the likelihood Equation \eqref{eq:likelihood_base}
can be written as 
\begin{equation}
\mathcal{L} (n_1,\dots,n_M, \gamma_1, \dots, \gamma_M)=
\prod_{i=1}^{N} \left[
\frac{1}{N} \sum_{k=1}^{M} n_k S^k_i(E_i, \delta_i, \alpha_i, \sigma_i, \gamma_k)
+ \frac{N-{n_s}}{N} B_i^{\mathrm{SigSub}}(E_i, \delta_i, \gamma_k)
\right]
\label{eq:likelihood_segmented}
\end{equation}
where $n_k=w_kn_s$ is the number of signal neutrinos observed from a given segment $k$.
The total number of signal neutrinos can then be obtained by summing the contribution 
of all segments $n_s=\sum_k n_k$.
The likelihood-ratio test statistic
\begin{equation}
\Lambda = 2 \ln \frac{\mathcal{L}(n_k=\hat n_k, \gamma_k = \hat\gamma_k)}{\mathcal{L}(n_s=0)}
\label{eq:ts}
\end{equation}
is defined by the ratio of the likelihood where all parameters $n_k, \gamma_k$ ($k=1,\dots, M)$ are optimized to maximize the likelihood,
compared to the null hypothesis of no neutrino emission $n_s=0$.

\subsection{Segmentation Schemes}
\begin{figure}[t!]
\centering

\begin{subfigure}{.49\textwidth}
\includegraphics[width=1\textwidth]{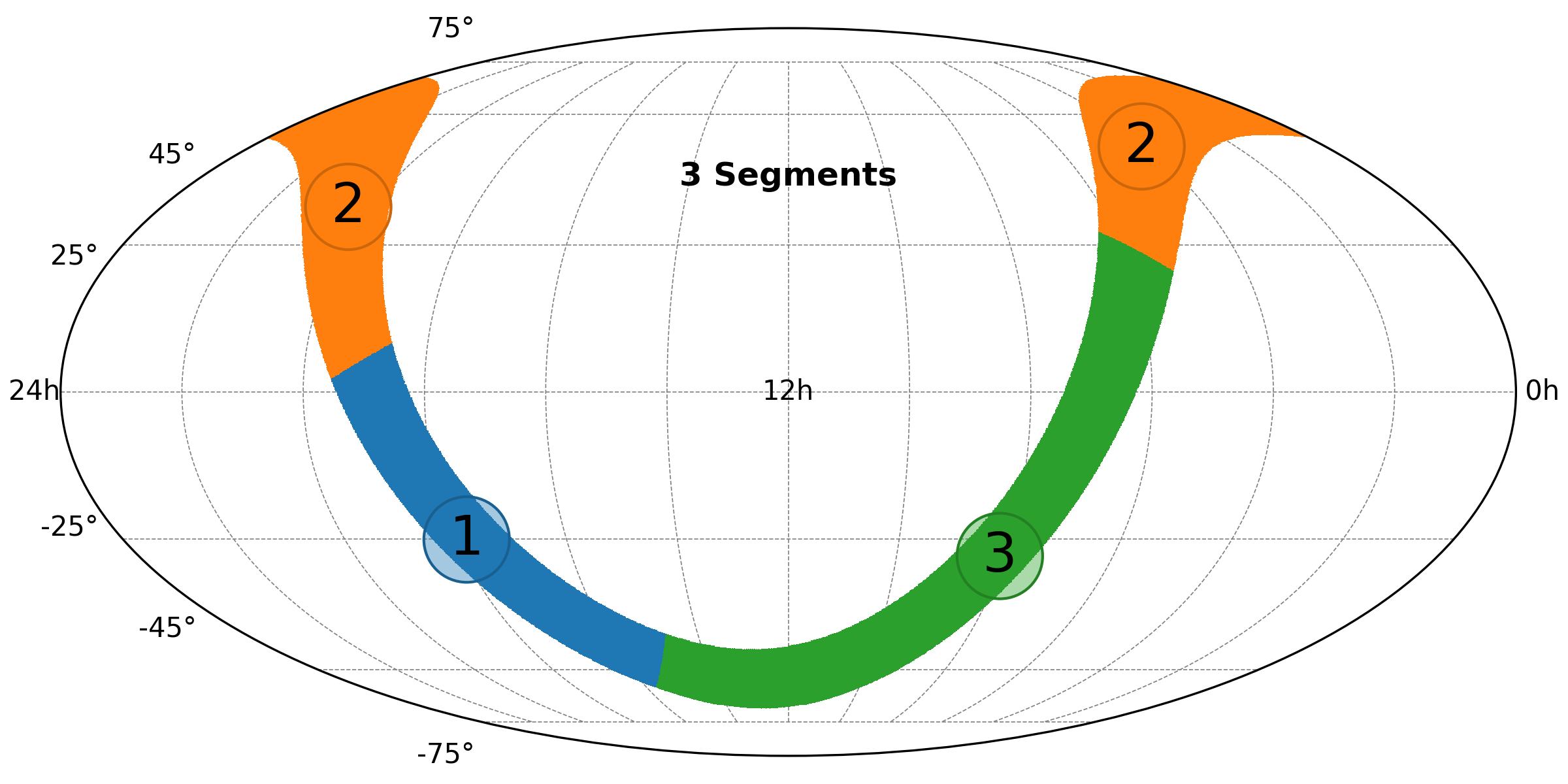}
\caption{The 3 segments \enquote{generic} segmentation scheme.}\label{fig:generic-3}
\end{subfigure}
\begin{subfigure}{.49\textwidth}
\includegraphics[width=1\textwidth]{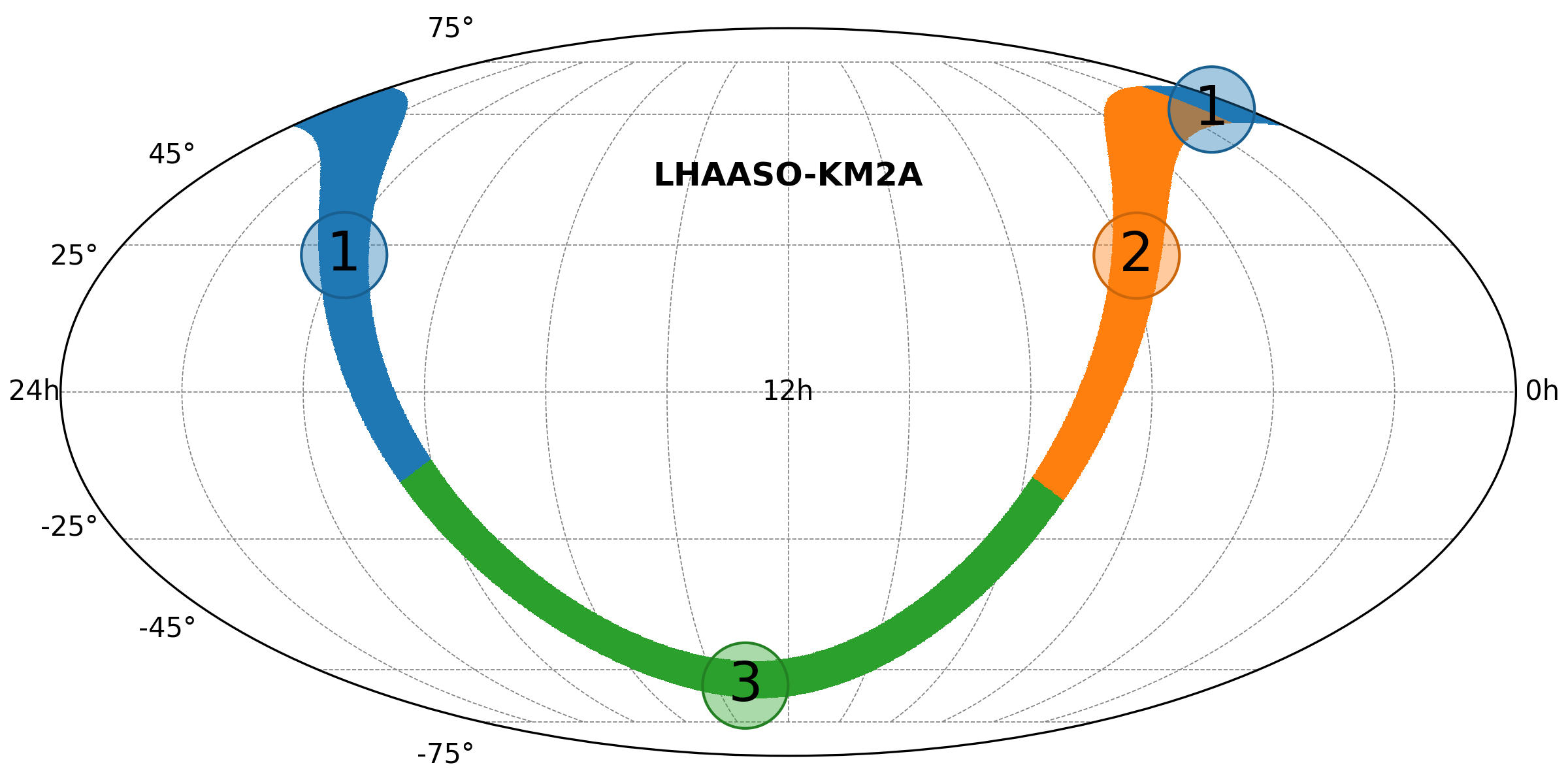}
\caption{Scheme aligning with the LHAASO GP analysis region.}\label{fig:lhaaso}
\end{subfigure}
\caption{Two of the eight used segmentation schemes of the galactic plan shown in equatorial coordinates in the Mollweide projection.}\label{fig:segmentation}

\end{figure}

Overall, eight different segmentation schemes are used to analyze the data.
This will result in eight test statistics.
First, three \enquote{generic} segmentation schemes are defined, 
which are chosen to obtain insights into the changing flux and spectrum 
at different galactic longitudes. 
All three generic segmentation schemes have segments that extend in galactic latitude from \qty{-8}{\degree} to \qty{8}{\degree},
which means they have a height of \qty{16}{\degree}.
This aligns with the inner galactic analysis region of the Fermi-LAT telescope in \rcite{Fermi-LAT:2012edv},
but is also a value which fits this analysis.
Since the cascades used in this analysis have an average angular uncertainty of \qty{7}{\degree} \cite{IceCube:2023ame},
choosing a smaller height than around \qty{14}{\degree} does not improve the sensitivity of this analysis. 
The \emph{first generic segmentation scheme} divides the galaxy into an inner region from 
galactic longitude \qtyrange{-60}{+60}{\degree} and \emph{one outer region} 
\qtyrange{-180}{-60}{\degree}, \qtyrange{60}{180}{\degree}.
The \emph{second generic segmentation scheme} divides the galaxy into an inner region 
from \qtyrange{-40}{+40}{\degree}, a left arm 
from \qtyrange{-180}{-40}{\degree} as well as a right 
arm from \qtyrange{40}{180}{\degree}.
This \enquote{3 Segments scheme} is visualized in \Cref{fig:generic-3} in
equatorial coordinates.
Then, to stress the analysis method, a segmentation scheme with 
six equal-size, adjacent segments is defined, each spanning a width in 
galactic longitude of \qty{60}{\degree}.
The central segment of that scheme is centered around the galactic center 
from \qty{-30}{\degree} to \qty{30}{\degree} in galactic longitude.
An overview of the different generic segmentation schemes along with their
unblinding results is provided in \Cref{tab:generic}.

In order to compare IceCube's GP results in a more model-independent way than previously possible,
five segmentation schemes aligning with analysis regions of high energy $\gamma$ ray observatories are defined.
These are HAWC \cite{Alfaro_2024}, H.E.S.S. \cite{HESS:2018pbp}, LHAASO\cite{LHAASO:2023gne, LHAASO:2024lnz} 
and Tibet AS$\gamma$ \cite{TibetASgamma:2021tpz}, an overview is provided in \Cref{tab:experiments}.
The two analysis regions of Tibet AS$\gamma$ overlap and must therefore be split 
into two different segmentation schemes. 
The chosen segmentation schemes are constructed by having the first segment(s) aligning 
with the analysis region(s) of the experiments and the last segment is
the remaining GP. This is illustrated in \Cref{fig:lhaaso} for the LHAASO segmentation scheme in equatorial coordinates.
There, Segment 3 covers the part of the GP which is not 
observed by LHAASO.
The obtained results can then be compared to a neutrino equivalent flux of the $\gamma$-fluxes.
Under simple assumptions they evaluate to
$E_\nu^2 \frac{\mathrm{d}N_\nu^{\text{All-Flavor}}}{\mathrm{d}E_\nu} = \sfrac{3}{2} E_\gamma^2 \frac{\mathrm{d}N_{\gamma}}{\mathrm{d}E_\gamma}$ and $E_\gamma = 2 E_\nu$ \cite{Fang:2023azx,Ahlers:2013xia}.

\section{Results}\label{sec2}

\begin{table}
    \centering
    \caption{Overview over the generic segmentation schemes. $l_{\mathrm{min}/\mathrm{max}}$ is the minimal/maximal galactic longitude of the given segment. 
            All segments are in between \qty{\pm 8}{\degree} of galactic latitude $b$. 
            The summed number of signal neutrinos $n_s$ from all segments and the significance of the analysis is also given.}\label{tab:generic}
    \begin{tblr}{
        colspec={l Q[si={table-format=1.0}] Q[si={table-format=+3.0}] Q[si={table-format=+3.0}] l l},
        row{1} = {guard},
        cell{2-12}{3-4} = {appto={\unit{\degree}}},
        cell{2,4,7}{5} = {appto={$\ \sigma$}},
    }
      \toprule
      Scheme     & Segment $k$ & $l_{\mathrm{min}}$ & $l_{\mathrm{max}}$ & Significance & $n_s=\sum_kn_k$ \\
      \midrule
      2 Segments & 1 & -60 & 60   & \SetCell[r=2]{l} $\left\} \rule{0pt}{2\ht\strutbox} \right.$ 3.81 & \SetCell[r=2]{l} 643  \\
                 & 2 & 60 & -60   &  &                        \\
      3 Segments & 1 & -40 & 40   & \SetCell[r=3]{l} $\left\} \rule{0pt}{3\ht\strutbox} \right.$ 3.84 & \SetCell[r=3]{l} 643 \\
                 & 2 & 40 & 180   &  &                        \\
                 & 3 & -180 & -40 &  &                        \\
      6 Segments & 1 & -30 & 30   & \SetCell[r=6]{l} $\left\} \rule{0pt}{5.9\ht\strutbox} \right.$ 3.58  & \SetCell[r=6]{l} 808 \\
                 & 2 & 30 & 90    &  &                        \\
                 & 3 & 90 & 150   &  &                       \\
                 & 4 & 150 & -150 &  &                        \\
                 & 5 & -150 & -90 &  &                       \\
                 & 6 & -90 & -30  &  &                       \\
      \bottomrule
    \end{tblr}
\end{table}

\begin{table}
    \centering
    \caption{Overview over the 5 segmentation schemes used to align with the galactic analysis regions of $\gamma$ ray telescopes. The region along the GP not covered by the $\gamma$ ray experiments is analysed as well, and corresponds to the latitude range not covered in the table. For LHAASO with one segmentation scheme two regions are analysed and both are provided in the table above each other.
    The significance of the scheme is given, as well as the sum of the best fit number of signal neutrinos from all segments $n_s$.
    The significance and number of neutrinos include also the segments not shown in the table and thus the whole range of galactic longitude.}
    \label{tab:experiments}
    \begin{tblr}{
    			colspec={l Q[si={table-format=+3.0}] Q[si={table-format=+3.0}] Q[si={table-format=1.0}] l Q[si={table-format=3.0}]},
    			row{1} = {guard},
    			cell{2-7}{2-4} = {appto={\unit{\degree}}},
    			cell{2-4,6-7}{5} = {appto={$\ \sigma$}},
    		}
    		\toprule
    		Experiment                                       & $l_{\min}$ & $l_{\max}$ & $\lvert{b}\rvert_{\mathrm{max}}$ & Significance  & $n_s=\sum_k n_k$ \\
    		\midrule
            HAWC \cite{Alfaro_2024}                          & 43         & 73         & 5         & 3.01 & 664 \\
            H.E.S.S \cite{HESS:2018pbp}                      & -110       & 65         & 3         & 3.06 & 569 \\
            LHAASO \cite{LHAASO:2023gne,LHAASO:2024lnz}      & 15         & 125        & 5         & \SetCell[r=2]{l} $\left\} \rule{0pt}{2\ht\strutbox} \right.$ 2.73 & \SetCell[r=2]{l} 371 \\
                                                             & 125        & -55        & 5         &  &  \\
            Tibet AS $\gamma$ I \cite{TibetASgamma:2021tpz}  & 25         & 100        & 5         & 2.96 & 660 \\
            Tibet AS $\gamma$ II \cite{TibetASgamma:2021tpz} & 50         & 200        & 5         & 2.98 & 626 \\
    		\bottomrule
    \end{tblr}
\end{table}

This analysis is designed to gain a deeper understanding of the previously reported 4.5$\sigma$ observation of the GP. Additional fit parameters are thus introduced, allowing for a more differentiated measurement at the cost of global significance. 

Despite the expected decrease in rejection power of the null hypothesis, the $p$-value can still be computed with the standard methodology 
of comparing the obtained test statistic $\Lambda$ Equation \eqref{eq:ts},
against the distribution of test statistics of background trials.
Background trials are constructed by randomizing the right-ascension $\alpha$
of each event. 
The unblinded significances along their best fit segment-wise number of signal neutrinos $n_k$ are presented in 
\Cref{tab:generic,tab:experiments}.
The most significant segmentation scheme turned out to be 
the one with 3 Segments with a value of $p = \num{6.1e-5} (3.84\sigma)$.
This result follows what is expected from this analysis method. 
When injecting simulation using the best fit value of the $\pi^0$ template from 
IceCube's previous GP template search \cite{IceCube:2023ame},
the median obtained significance is $3.57\sigma$ for this analysis.

The background distribution of $\Lambda$ for each segmentation 
scheme is very different from one-another, due to the different number of 
free parameters.
In order to obtain a global $p$-value of all analyses, a trials 
correction must be performed.
This is done with correlated background trials, in which for one given 
background scramble all 8 analyses are performed and the smallest $p$-value 
among them is saved.
The smallest unblinded $p$-value can then be compared to the distribution 
of smallest $p$-values on background trials.
For this analysis this gives an overall value of $p=\num{2.97e-4} (3.43\sigma)$,
which is a factor of 4.9 higher than the best local $p$-value.
This is better than applying the Šidák correction \cite{Sidak01061967}, which would result 
in a trials factor of $(1-(1-p)^8)\approx 8 p$.
It shows that the analyses are, as expected, highly correlated. 

\subsection{Likelihood Contours}
\begin{figure}
\centering
\includegraphics[width=.6\textwidth]{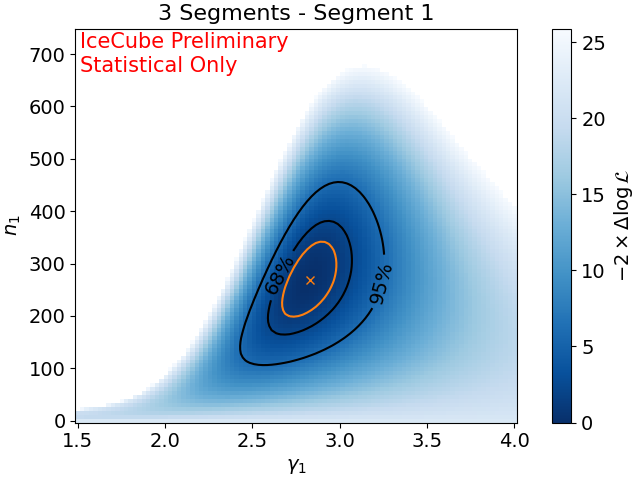}
\caption{
The $n_k$-$\gamma_k$ likelihood contour of the inner galaxy ($\lvert l \rvert < \qty{40}{\degree}$, $\lvert b \rvert < \qty{8}{\degree}$ \enquote{Segment 1}) in the 3-segments generic segmentation scheme, visualized in blue in \Cref{fig:generic-3}. 
The black \qty{68}{\percent} and \qty{95}{\percent} contours are drawn assuming Wilks' theorem assuming 2 degrees of freedom, 
the orange \qty{68}{\percent} contour is drawn assuming Wilks' theorem with 1 degree of freedom. The orange cross is the best fit point. The shown contours only show statistical uncertainty, no systematics treatment is included here.
}\label{fig:contour_inner}
\end{figure}

\begin{figure}
\begin{subfigure}{0.49\textwidth}
\includegraphics[width=\textwidth]{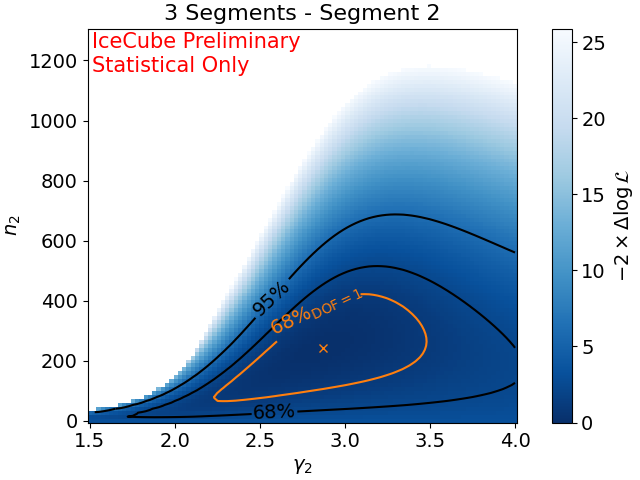}
\end{subfigure}
\begin{subfigure}{0.49\textwidth}
\includegraphics[width=\textwidth]{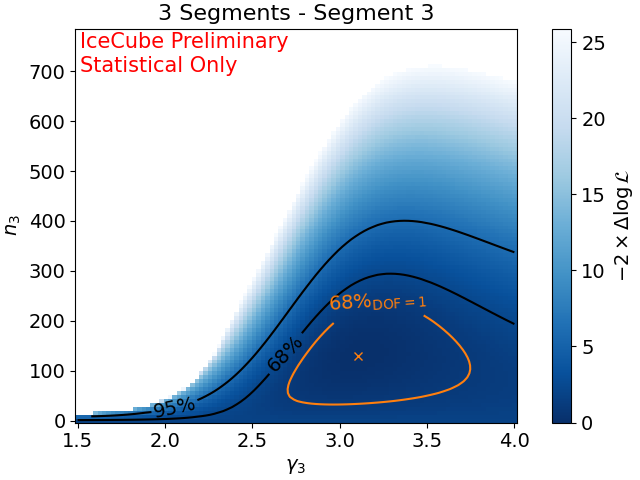}
\end{subfigure}
\caption{
The $n_k$-$\gamma_k$ likelihood contours of the left and right arm of the galaxy (Segment 2 and 3) in the 3-segments generic segmentation scheme, visualized in \Cref{fig:generic-3}. 
The black \qty{68}{\percent} and \qty{95}{\percent} contours are drawn assuming Wilks' theorem assuming 2 degrees of freedom, 
the orange \qty{68}{\percent} contour is drawn assuming Wilks' theorem with 1 degree of freedom. The orange cross is the best fit point. The shown contours only show statistical uncertainty, no systematics treatment is included here.
}\label{fig:contours}
\end{figure}

To obtain uncertainty intervals or contours on the obtained parameters the profile likelihood method can be used.
In practice for this analysis, it means that one can evaluate the parameters $n_k$ and $\gamma_k$ of one segment through a grid of points,
while re-maximizing the likelihood by varying all other parameters.
Since the segments are spatially disjunct and large compared to the angular uncertainty, the segments are nearly perfectly uncorrelated 
among each other, so profiling over the other segments makes little difference.
The likelihood contour for the inner galaxy is shown in \Cref{fig:contour_inner}  and the two contours 
for the arms of the galaxy are shown in
\Cref{fig:contours}.
The confidence contours are drawn by assuming Wilks' theorem \cite{Wilks:1938dza}.
For convenience, the contours assuming one degree of freedom in the $\chi^2$ distribution 
are shown in orange, which enables a direct extraction of the profiled uncertainties per-parameter.

\section{Conclusion}\label{sec3}
This work presents a first application of a new method to characterize neutrino emission from the GP.
Applied to the same dataset used in the original observation of the GP \cite{IceCube:2023ame}, 
it can provide confidence regions of the flux normalization and the spectral index of a simple unbroken power law,
as shown in \Cref{fig:contour_inner,fig:contours}.
The first measurement of the spectral index $\gamma$ of the neutrino flux in the inner galaxy is shown in \Cref{fig:contour_inner}.

The presented method does not rely on a fine-grained template of the spatial distribution of neutrino emission 
from the GP, but rather makes the simple choice of assuming uniform emission in each segment.
While the outer regions yield larger contours, they are expected to shrink with updated datasets.

In the future, this analysis can be applied to updated datasets using updated reconstructions, combined event topologies and
more data.
The data-driven estimation of the background distribution has the advantage that it does not rely on 
modeling of the atmospheric neutrino flux, which comes with a large uncertainty.
Measuring a spectral index of the galactic neutrino emission enables to enhance the comparison 
between gamma ray prediction or direct results and in turn helps understanding the origin
of the galactic neutrino flux better.
It is a crucial step in differentiating hadronic and leptonic emission from gamma rays
and allows estimating contributions from neutrino sources in the GP.

\FloatBarrier
\bibliographystyle{ICRC}
\bibliography{references}

%

\clearpage

\section*{Full Author List: IceCube Collaboration}

\scriptsize
\noindent
R. Abbasi$^{16}$,
M. Ackermann$^{63}$,
J. Adams$^{17}$,
S. K. Agarwalla$^{39,\: {\rm a}}$,
J. A. Aguilar$^{10}$,
M. Ahlers$^{21}$,
J.M. Alameddine$^{22}$,
S. Ali$^{35}$,
N. M. Amin$^{43}$,
K. Andeen$^{41}$
% ...
\\
\\
% ...
$^{10}$ Universit{\'e} Libre de Bruxelles, Science Faculty CP230, B-1050 Brussels, Belgium \\
% ...
$^{16}$ Department of Physics, Loyola University Chicago, Chicago, IL 60660, USA \\
$^{17}$ Dept. of Physics and Astronomy, University of Canterbury, Private Bag 4800, Christchurch, New Zealand \\
% ...
$^{21}$ Niels Bohr Institute, University of Copenhagen, DK-2100 Copenhagen, Denmark \\
$^{22}$ Dept. of Physics, TU Dortmund University, D-44221 Dortmund, Germany \\
% ...
$^{35}$ Dept. of Physics and Astronomy, University of Kansas, Lawrence, KS 66045, USA \\
% ...
$^{39}$ Dept. of Physics and Wisconsin IceCube Particle Astrophysics Center, University of Wisconsin{\textemdash}Madison, Madison, WI 53706, USA \\
% ...
$^{41}$ Department of Physics, Marquette University, Milwaukee, WI 53201, USA \\
% ...
$^{43}$ Bartol Research Institute and Dept. of Physics and Astronomy, University of Delaware, Newark, DE 19716, USA \\
% ...
$^{63}$ Deutsches Elektronen-Synchrotron DESY, Platanenallee 6, D-15738 Zeuthen, Germany \\
$^{\rm a}$ also at Institute of Physics, Sachivalaya Marg, Sainik School Post, Bhubaneswar 751005, India \\
% ...

\subsection*{Acknowledgments}

\noindent
The authors gratefully acknowledge the support from the following agencies and institutions:
USA {\textendash} U.S. National Science Foundation-Office of Polar Programs,
U.S. National Science Foundation-Physics Division,
U.S. National Science Foundation-EPSCoR,
U.S. National Science Foundation-Office of Advanced Cyberinfrastructure,
Wisconsin Alumni Research Foundation,
Center for High Throughput Computing (CHTC) at the University of Wisconsin{\textendash}Madison,
Open Science Grid (OSG),
Partnership to Advance Throughput Computing (PATh),
Advanced Cyberinfrastructure Coordination Ecosystem: Services {\&} Support (ACCESS),
Frontera and Ranch computing project at the Texas Advanced Computing Center,
U.S. Department of Energy-National Energy Research Scientific Computing Center,
Particle astrophysics research computing center at the University of Maryland,
Institute for Cyber-Enabled Research at Michigan State University,
Astroparticle physics computational facility at Marquette University,
NVIDIA Corporation,
and Google Cloud Platform;
Belgium {\textendash} Funds for Scientific Research (FRS-FNRS and FWO),
FWO Odysseus and Big Science programmes,
and Belgian Federal Science Policy Office (Belspo);
Germany {\textendash} Bundesministerium f{\"u}r Forschung, Technologie und Raumfahrt (BMFTR),
Deutsche Forschungsgemeinschaft (DFG),
Helmholtz Alliance for Astroparticle Physics (HAP),
Initiative and Networking Fund of the Helmholtz Association,
Deutsches Elektronen Synchrotron (DESY),
and High Performance Computing cluster of the RWTH Aachen;
Sweden {\textendash} Swedish Research Council,
Swedish Polar Research Secretariat,
Swedish National Infrastructure for Computing (SNIC),
and Knut and Alice Wallenberg Foundation;
European Union {\textendash} EGI Advanced Computing for research;
Australia {\textendash} Australian Research Council;
Canada {\textendash} Natural Sciences and Engineering Research Council of Canada,
Calcul Qu{\'e}bec, Compute Ontario, Canada Foundation for Innovation, WestGrid, and Digital Research Alliance of Canada;
Denmark {\textendash} Villum Fonden, Carlsberg Foundation, and European Commission;
New Zealand {\textendash} Marsden Fund;
Japan {\textendash} Japan Society for Promotion of Science (JSPS)
and Institute for Global Prominent Research (IGPR) of Chiba University;
Korea {\textendash} National Research Foundation of Korea (NRF);
Switzerland {\textendash} Swiss National Science Foundation (SNSF).

\end{document}